\documentclass[12pt]{article}

\usepackage{scicite}

\usepackage{amsmath} 
\usepackage{times}
\usepackage{graphicx}
\usepackage{color,soul}
\usepackage{tikz}
\newcommand*\circled[1]{\tikz[baseline=(char.base)]{
            \node[shape=circle,draw,inner sep=1pt] (char) {#1};}}

\newenvironment{sciabstract}{%
\begin{quote} \bf}
{\end{quote}}

\newcounter{lastnote}

\usepackage{comment}

\title{Laser-power consumption of soliton formation in a bidirectional Kerr resonator} 
\author
{Jizhao Zang$^{1,2\ast}$, Su-Peng Yu$^{1,2}$, Haixin Liu$^{1,2}$, Yan Jin$^{1,2}$, Travis C. Briles$^{1}$,\\David R. Carlson$^{1,3}$,
and Scott B. Papp$^{1,2}$\\
\\
\normalsize{$^{1}$Time and Frequency Division, National Institute of Standards and Technology, Boulder, CO, USA}\\
\normalsize{$^{2}$Department of Physics, University of Colorado, Boulder, CO, USA}\\
\normalsize{$^{2}$Octave Photonics, Louisville, CO, USA}\\
\\
\normalsize{$^\ast$E-mail:  jizhao.zang@nist.gov.}
}

\date{}

\begin{document} 

\maketitle 

\begin{sciabstract}
Laser sources power extreme data transmission as well as computing acceleration, access to ultrahigh-speed signaling, and sensing for chemicals, distance, and pattern recognition. The ever-growing scale of these applications drives innovation in multi-wavelength lasers for massively parallel processing. We report a nanophotonic Kerr-resonator circuit that consumes the power of an input laser and generates a soliton frequency comb at approaching unit efficiency. By coupling forward and backward propagation, we realize a bidirectional Kerr resonator that supports universal phase matching but also opens excess loss by double-sided emission. Therefore, we induce reflection of the resonator's forward, external-coupling port to favor backward propagation, resulting in efficient, one-sided soliton formation. Coherent backscattering with nanophotonics provides the control to put arbitrary phase-matching and efficient laser-power consumption on equal footing in Kerr resonators. In the overcoupled-resonator regime, we measure 65\% conversion efficiency of a 40 mW input pump laser, and the nonlinear circuit consumes 97\% of the pump, generating the maximum possible comb power. Our work opens up high-efficiency soliton formation in integrated photonics, exploring how energy flows in nonlinear circuits and enabling laser sources for advanced transmission, computing, quantum sensing, and artificial-intelligence applications.
\end{sciabstract}

\section*{}
Nonlinear optics is a ubiquitous phenomenon in science and technology that converts light from one wavelength to another and enables the synthesis of new optical fields. Development in nonlinear optics has advanced a host of applications, ranging from optical signal processing \cite{Willner2014} to ultra-short pulsed lasers \cite{David2018} and optical parametric amplifiers \cite{Hansryd2002,Foster2006,Riemensberger2022} to quantum-state generation and manipulation \cite{Avik2015,Morandotti2016,Xu2021NC,Jelena2022,Xu2023Optica}. Invariably, nonlinear effects are relatively weak, and harnessing their benefits stimulates discovery and advancements in engineering the control of light-matter interactions. Since the third-order Kerr-effect nonlinearity involving intensity-dependent refractive index occurs in practically all materials, it is widely applied in scalable technologies like integrated photonics \cite{TobiasK2011,Moss2013,Morandotti2018,TobiasK2018,Gaeta2019}.


Controlling Kerr nonlinearity in microresonators is a central goal for integrated photonics. The high quality factor and small mode volume of microresonators make optical parametric oscillation possible with a threshold power much less than one milliwatt \cite{Ji2017}, enabling the use with many different types of lasers. Moreover, adjusting the material and shape of a microresonator enables phase-matching for four-wave mixing (FWM). There now exists a fundamental understanding of stability criteria for many types of nonlinear field states in Kerr microresonators, including parametric oscillation \cite{Kartik2020,JenBlack2022,Kartik2023} and soliton microcombs \cite{TobiasHerr2014,Yi&Vahala2015,TobiasK2018,Erkintalo2021}. We consider the latter as self-reinforcing, temporally localized pulses in a resonator that manifest as a frequency comb at the resonator output. 
Recently, soliton stability and generation has been enhanced by coupling forward and backward propagation in microresonators as an added mechanism to control nonlinear phase matching. Photonic-crystal resonator (PhCR) microcombs \cite{Su-peng2021,Su-peng2022,Erwan2023} implement coherent scattering with sub-wavelength nanopatterns to realize a bidirectional Kerr resonator, and self-injection-locked microcombs exploit direct laser-resonator feedback \cite{Vahala&Bowers2020,TobiasK&Igor2021,Tobias&Bowers2021,Briles2021,TobiasHerr2024}. These bidirectional systems can feature universal phase matching in any dispersion regime and spontaneous soliton formation.

While phase-matching is the essential condition to initiate nonlinear field states like soliton microcombs, efficiently energizing and harvesting the soliton comb represents a distinct set of physical optimizations. We define the pump-to-comb conversion efficiency (CE) as $\rm{CE}=P_{\rm{comb}}/P_{\rm{pump}}$, where $P_{\rm{pump}}$ is the pump laser power and $P_{\rm{comb}}$ is the total, emitted comb power. High CE increases the power per comb mode, which is critical for applications like wavelength-division-multiplexing (WDM) transceivers \cite{Kiyoul2022,Bergman2023} and microwave photonic filters \cite{DMoss2020}. To maximize CE, first, the external coupling rate ($\kappa_c$) of the microresonator must be larger than the internal loss rate ($\kappa_i$). Furthermore, phase-matching constrains the CE of solitons because only limited ranges of system parameters like laser detuning are allowed. Still, isolated, high CE soliton states have been reported. Efficient temporal solitons have been demonstrated in coupled fiber cavities \cite{Xue2019} and a microresonator nested in a fiber loop with gain \cite{Moss&Pasquazi2019}, but those systems are not compatible with chip-scale integration. Soliton crystals achieve nearly unit CE by exciting several solitons in one resonator, however, destructive interference reduces the utility of such ensembles \cite{DanCole2017}. Solitons in anomalous-dispersion resonators with artificially shifted pump resonance offer a path to high CE \cite{Victor2023}, but precise control of the pump laser and auxiliary cavity is required to initiate the soliton state. Solitons in normal-dispersion resonators \cite{Xue2015,Xue2017,Gaeta2019highCE,Victor2021} can leverage a long duty cycle for enhanced CE, however soliton phase-matching is operationally challenging to achieve. Hence, unified control of phase-matching and CE for high pump-laser consumption in solitons is an outstanding objective.

Here, we leverage coherent scattering in a nanophotonic Kerr-resonator circuit to provide universal phase-matching for soliton microcombs and demonstrate deterministic control of CE up to the unit boundary. We experimentally generate solitons with a PhCR that features normal dispersion and one 'split mode' with coherently coupled forward and backward propagation. Moreover, this direct phase matching enables a continuous evolution from the empty resonator state to a backward propagating soliton. However, CE in a bidirectional resonator cannot exceed 50\% because the localized soliton circulates in one direction while the pump is distributed in both. To consume the forward pump power and close off excess loss of the double-sided resonator, we place a reflector on the coupling waveguide. Our nanophotonic soliton circuit emits a 200 GHz train of flattop pulses with up to 65\% CE and 97\% pump consumption. Moreover, the nanophotonic circuit enables precise control of the comb spectrum that we use for alignment to the ITU-T grid with $<2.6$ GHz offset and compatibility with a distributed feedback (DFB) laser diode on the 193.1 THz channel.  Our work demonstrates near unit efficiency comb generation with a bidirectional Kerr resonator by controlling pump utilization with on-chip nanophotonics, enabling for example massive scaling in data transmission and processing.

 \begin{figure}[htb]
\centering
\includegraphics[width=1\linewidth,trim={00cm 0.1cm 00cm 0cm},clip]{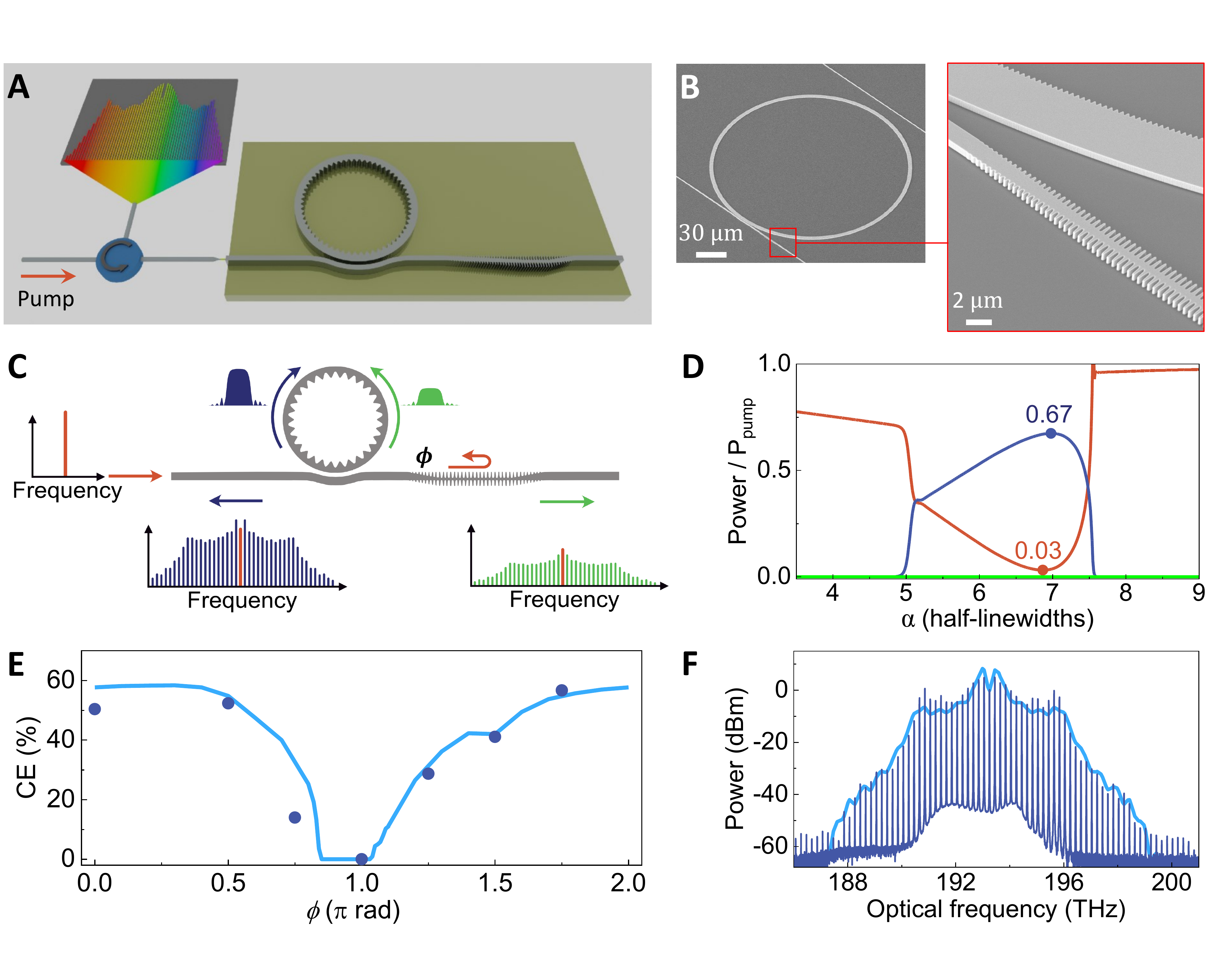}
\caption{Soliton formation in the nanophotonic circuit. (A) Schematic of the nanophotonic Kerr resonator circuit with a PhCR and a waveguide reflector. (B) Scanning electron microscope images of the nanophotonic structures. (C) The bidirectional pump in the PhCR supports excitation of solitons in the forward and backward directions. We tune $\phi$ for constructive interference between the reflected pump and the backward pump in the PhCR. (D) Power flow in the circuit analyzed by LLE simulation with $(P_f+P_b)/P_{\rm{pump}}$ (red), $P_{\rm{comb}}^{f}/P_{\rm{pump}}$ (green), and $P_{\rm{comb}}^b/P_{\rm{pump}}$ (blue). (E) Measured CE versus $\phi$ (circles) and LLE simulation (line). (F) Measured (data trace) and simulated (line) optical spectrum of a soliton microcomb with a CE of 65\%.   }
\label{Fig1}
\end{figure}

Figure 1 introduces the structure of our nanophotonic resonator circuit with an emphasis on how we achieve unified control of phase-matching and CE through coherent backscattering. The circuit (Fig. 1A) is composed of: a PhCR in which the ring width (RW) is modulated with a periodicity of $\pi a/m$, where $a$ is the PhCR radius and $m$ is the azimuthal mode number; an evanescent coupling waveguide curved around the PhCR for enhanced escape efficiency; and a reflector in the coupling waveguide in which the waveguide width is modulated with a periodicity of $\Lambda=\pi c /\omega_l n_{\rm{eff}}$, where c is the speed of light,  $\omega_l$ is the laser frequency, and $n_{\rm{eff}}$ is the effective refractive index. We denote PhCR mode $m$ as the split mode, which manifests as two resonances split in frequency by the bandgap $\epsilon$ due to the coherently coupled forward and backward propagation.  To energize a soliton in a normal dispersion PhCR, we tune the pump laser into the lower frequency resonance of the split mode, which offers optimal phase-matching.  The higher frequency resonance of the split mode does not offer phase-matching for FWM because the adjacent PhCR modes are all at lower frequencies. Since we induce nearly complete reflection over a 3 dB bandwidth of 7.6 THz, the forward-propagating pump laser and backward-propagating soliton microcomb combine at a circulator and couple to the chip at a common facet. We implement the circuit on a silicon chip by use of conventional semiconductor processing with the tantalum pentoxide ($\rm{Ta_2O_5}$, hereafter tantala) material platform. We perform nanofabrication on a 75 mm wafer with $<$150 nm feature resolution in the tantala device layer, creating 50 chips with 100 circuit variations per chip for experiments \cite{Lamee2020,Jung2021,Black2021,Su-peng2021}. Using electron-beam lithography, we are able to precisely control the circuit dimensions, especially RW modulation at the nanometer scale. Figure 1B shows scanning electron microscope images of the nanophotonic circuit elements, emphasizing the modulation on the inner edge of the PhCR and the width-modulated waveguide that forms the reflector. 

To understand operation of the nanophotonic circuit, we consider the forward and backward propagating fields; see Fig. 1C. Coherent scattering in the PhCR creates a bidirectional pump that in principle supports simultaneous excitation of solitons in the forward and backward directions. However, backward operation shows higher gain and is favored over forward operation \cite{Erwan2023}. Hence, the pump laser in the forward direction is not used. The function of the reflector is to reflect the forward pump field and constructively interfere with the backward pump field in the PhCR. We characterize the phase delay $\phi$ of the reflected pump with respect to the PhCR, and operationally, we control $\phi$ by rotating the PhCR with respect to the coupling waveguide.  

We explore the dynamics and steady-state field solution of the PhCR with the Lugiato-Lefever equation (LLE) \cite{LLE1987}, including the forward-backward coupling and the pump reflector that we implement through coherent scattering \cite{Haixin2024}. Moreover, the LLE depends on dissipation, group-velocity dispersion (GVD), and detuning of the pump laser. In particular, 
\begin{equation} 
\begin{split}
    \frac{\partial E^f_{\mu}}{\partial t} = -(1+i \alpha)E^f_{\mu}-i \frac{d_{2}}{2} \mu^2 E^f_{\mu}+i \mathcal{F}(|E^{f}|^2E^{f})_\mu+2i E^f_{\mu} \sum_j|E^b_j|^2  \\+\delta_{\mu0}(F-\frac{i\epsilon_{\rm{PhC}}}{2}E^b_{\mu}) \\
 \frac{\partial E^b_{\mu}}{\partial t} = -(1+i \alpha)E^b_{\mu}-i \frac{d_{2}}{2} \mu^2 E^b_{\mu}+i \mathcal{F}(|E^{b}|^2E^{b})_\mu+2i E^b_{\mu} \sum_j|E^f_j|^2 \\+\delta_{\mu0}(rF-\frac{i\epsilon_{\rm{PhC}}}{2}E^f_{\mu}) -I_\Omega(\mu)\gamma r E^f_{\mu}
\end{split}
\end{equation}
where $E^f_{\mu}$ and $E^b_{\mu}$ are the forward and backward field amplitude in mode with index $\mu$ relative to the split mode. All frequency variables are normalized to the half-linewidth of the resonator $\kappa/4\pi$, where $\kappa=\kappa_i+\kappa_c$ is the total loss rate. The variable $\alpha=2(\omega_l-\omega_0)/\kappa$ is the normalized detuning between $\omega_l$ and the split mode $\omega_0$. The parameter $d_2$ is the normalized GVD $D_2$ that is derived from the Taylor expansion of the resonance frequencies around $\omega_0$:  $\omega_\mu=\omega_0+D_1\mu+D_2\mu^2/2+\cdots$, where $D_1 /2\pi$ is the free-spectral range (FSR).  The third and fourth terms of Equation (1) stand for self-phase modulation and cross-phase modulation, respectively, where $E^f$ and $E^b$ are the temporal amplitudes of the intraresonator field and $\mathcal{F}()_\mu$ represents the $\mu$th frequency component of the Fourier series.  Different sources contribute to the  pump power in the forward and backward directions: in the forward direction, the pump power originates from the external laser on the coupling waveguide and the power coupled from $E^b_{0}$ due to coherent backscattering in the PhCR; in the backward direction, the pump power originates from the external laser after travelling through the evanescent coupling region and being reflected, $E^f_{0}$ is outcoupled, reflected and coupled back into the PhCR,  and coherent backscattering of $E^f_{0}$ inside the resonator.  In the fifth term,  $\delta_{\mu0}$ is the Kronecker delta ($\delta_{00}=1$, otherwise $\delta_{\mu0}=0$), $r=\sqrt{R}\ e^{i\phi}$ is the reflection coefficient of the waveguide reflector ($R$ is the reflectivity), $F$ is the normalized amplitude of the pump, and $\epsilon_{\rm{PhC}}=4\pi\epsilon/\kappa$ is the normalized $\epsilon$. The transfer function from the resonator to the coupling waveguide is defined as $\gamma=2K/(K+1)$, where $K$ is the coupling factor. Since the reflector has a finite reflection bandwidth, we apply an indicator function $I_\Omega(\mu)$: if $\mu\in\Omega$, $I_\Omega(\mu)=1$, otherwise $I_\Omega(\mu)=0$, where $\Omega$ is a set of $\mu$ that corresponds to the resonator modes in the reflection band. In LLE simulation, we use $\Omega=\{-19,-18,\ldots,18,19\}$ to model our nanophotonic circuit and we assume $r$ is the same for $\mu \in \Omega$.

Using the LLE, we simulate power flow in the nanophotonic circuit to understand CE and pump laser consumption; see Fig. 1D. To quantify the power flow in both propagating directions on the coupling waveguide, we denote the comb power and the outcoupled pump power in the forward and backward directions as $P_{\rm{comb}}^{f}$, $P_{\rm{comb}}^{b}$, $P_{f}$ and $P_{b}$, respectively. Then, $P_{\rm{comb}}$ naturally becomes $P_{\rm{comb}}^{f}+P_{\rm{comb}}^{b}$. As a function of $\alpha$, we simulate $(P_{f}+P_{b})/P_{\rm{pump}}$ (red), $P_{\rm{comb}}^{f}/P_{\rm{pump}}$ (green), and $P_{\rm{comb}}^{b}/P_{\rm{pump}}$ (blue) in the coupling waveguide. Since the larger gain and the reflector favor backward soliton formation, we consider $P_{\rm{comb}}^{b} \gg P_{\rm{comb}}^{f}$ and $P_{\rm{comb}} \approx P_{\rm{comb}}^{b}$ unless otherwise specified. In the simulation, the optimal reflection phase for maximum CE is $\phi=0$ rad. The other parameter of this simulation is $K$, which we set to a value of 4.5 based on the likelihood of successfully fabricating a coupler with this value. As we inspect the simulated behavior of soliton formation in this device in a similar fashion as we sweep $\alpha$ in the experiment, we observe initial consumption of the pump laser due to internal losses of the resonator. Then we observe rapid formation of the soliton, reaching CE and laser consumption of 67\% and 97\%, respectively. As $\alpha$ exceeds the range over which the soliton is stable in the resonator, the soliton rapidly de-energizes. These predicted behaviors are favorable for applications, especially the high CE, hence we are motivated to explore soliton formation with fabricated versions of this system.  

Our simulations indicate that $\phi$ is the critical parameter to access the regime of near unit CE in our nanophotonic circuit. Hence, we design a set of circuits on a common chip in which we systematically and precisely vary $\phi$; by testing such copies of the circuit, we search for optimum CE. Here, our test procedure is to scan $\alpha$, record the reflection of the split mode to characterize $\phi$ dependence of pump interference, and then pump the lower frequency resonance of the split mode with optimal $\phi$.

Figure 1E presents measurements of CE (circles) in the nanophotonic circuit alongside the LLE simulation (line) as a function of $\phi$. The agreement of the simulation and experiment highlights our understanding of the system, our capability to accurately fabricate specific circuit designs with a set of $\phi$ values, and robustness of the nonlinear nanophotonic circuit operation. The variation in $\phi$ results in constructive and destructive interference between the reflected forward pump by the reflector and backward pump in the PhCR, leading to different CE in the nanophotonic circuit. Here, we use $K=3$ and the highest CE is 57\% in the experimental data, but our simulation suggests that higher CE is possible with a larger $K$. For example, we observe a CE of 65\% at $K=4.5$ and $P_{\rm{pump}}=40$ mW. In Fig. 1F, we present the optical frequency spectrum that we obtain with this highest CE device. We tune $\alpha$ to access this spectrum in which the mode-to-mode power variation is relatively small. The measured soliton microcomb spectrum (data trace) is in agreement with the LLE simulation (line). We also observe substantial consumption of the pump laser in forming solitons, and the remaining pump power exiting the circuit is only 1 mW.

 \begin{figure}[htb]
\centering
\includegraphics[width=1\linewidth,trim={0cm 4.4cm 0cm 1.1cm},clip]{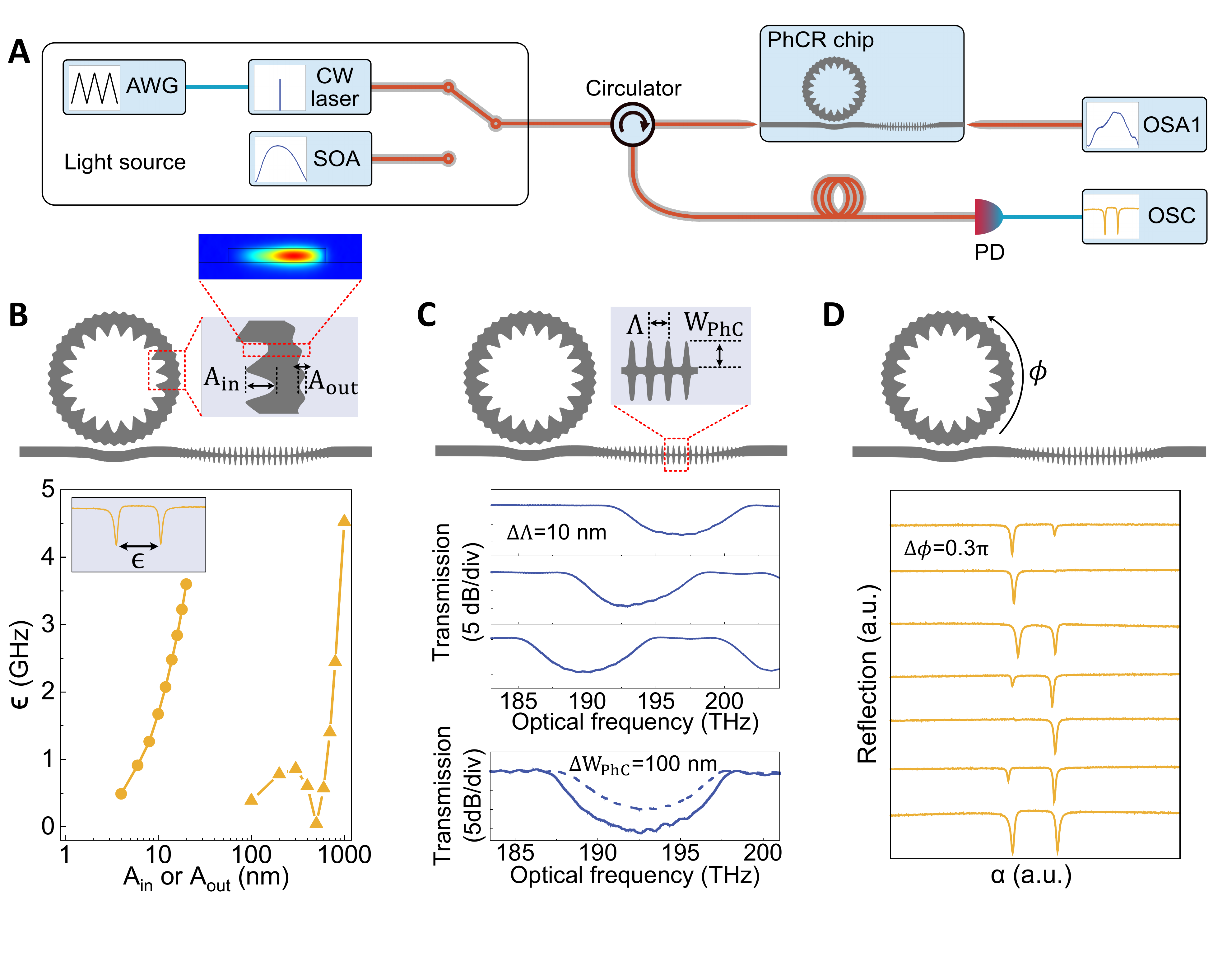}
\caption{ Characterization of the nanophotonic circuit structures. (A) Setup for device characterization. AWG, arbitrary waveform generator; CW laser, continuous-wave laser; SOA, semiconductor optical amplifier; OSA, optical spectrum analyzer; PD, photodetector; OSC, oscilloscope. (B) Measured $\epsilon$ as a function of the modulation amplitude on the inner ($A_{\rm{in}}$, triangles) or outer ($A_{\rm{out}}$, squares) edge of the PhCR ($a=109.5$ $\mu$m, RW = 4 $\mu$m). Inset: transmission of a split mode. (C) Optimization of the reflection band and $R$: transmission spectra of the reflector for varied $\Lambda$ and $W_{\rm{PhC}}$. (D) Nanophotonic circuit reflection as we step $\phi$ in amounts of $0.3\pi$. }
\label{Fig2}
\end{figure}

We present characterization of the nanophotonic circuit structures that control CE; see Fig. 2. We use a narrow linewidth laser and a broadband light source to measure with fine frequency resolution and across the entire operating bandwidth of the circuit, respectively. We record optical signals with a standard optical spectrum analyzer and a photodetector with an emphasis on the reflected power from the circuit; see Fig. 2A. 

As we show in Fig. 2B, we implement the azimuthal RW modulation ($A_{\rm{in}}$ and $A_{\rm{out}}$) on either the inner or outer edge of the PhCR, since the overlap between the interactivity field and the nanopattern permits access to different ranges of modulation amplitude. With the PhCR, the number of modulation periods corresponds to twice $m$, the mode number of the PhCR mode that exhibits coherent backscattering, and the amplitude of the modulation controls $\epsilon$ of the split mode. We measure $\epsilon$ by a calibrated laser frequency scan across the split mode, and access to $\epsilon$ from zero to 5 GHz provides the range that our LLE simulations indicate are needed for soliton formation. To implement and test the reflector in the coupling waveguide, we tune $\Lambda$ so that the reflection band is centered on the split mode, and we optimize the amplitude of the nanopattern for $R>90\%$. Figure 2C shows our operational procedure to optimize the reflection band as we vary $\Lambda$ to shift the frequency of maximum reflection. We observe this shift in the transmission spectra as we step $\Lambda$ in steps $\Delta\Lambda=10$ nm. This relatively coarse resolution in the nanopattern is sufficient to align the split mode and peak $R$ of the reflector. We also characterize the dependence of $R$ on the amplitude of the reflector nanopattern; see the modulation depth $W_{\rm{PhC}}$ in Fig. 2C. By adjusting $W_{\rm{PhC}}$ from 750 nm to 850 nm, we vary $R$ from 90\% to 97\%.

With optimized reflection bandwidth and $R$, we characterize the intraresonator pump power at the two resonances of the split mode. Figure 2D shows measurements of reflection from the nanophotonic circuit in which we step $\phi$ in amounts of $0.3\,\pi$. The data demonstrates interference of the pump laser, particularly destructive interference of either the higher or lower frequency component of the split mode. Due to the electric field profile in the PhCR, the two resonances are out of phase by $\pi$ \cite{Su-peng2021}. Operationally, we vary $\phi$ by rotating the PhCR to change the alignment of the azimuthal nanopattern and the reflector. This procedure mitigates fabrication imperfections due to the rotational symmetry of the resonator. Another benefit of the reflector in our nanophotonic circuit is reduction in thermal instability from absorbed pump power in the resonator. At the optimal setting of $\phi$, the higher frequency resonance is not substantially excited and hence does not contribute to absorption and heating of the PhCR.

 \begin{figure}[htb]
\centering
\includegraphics[width=1\linewidth,trim={1.2cm 2cm 00cm 0cm},clip]{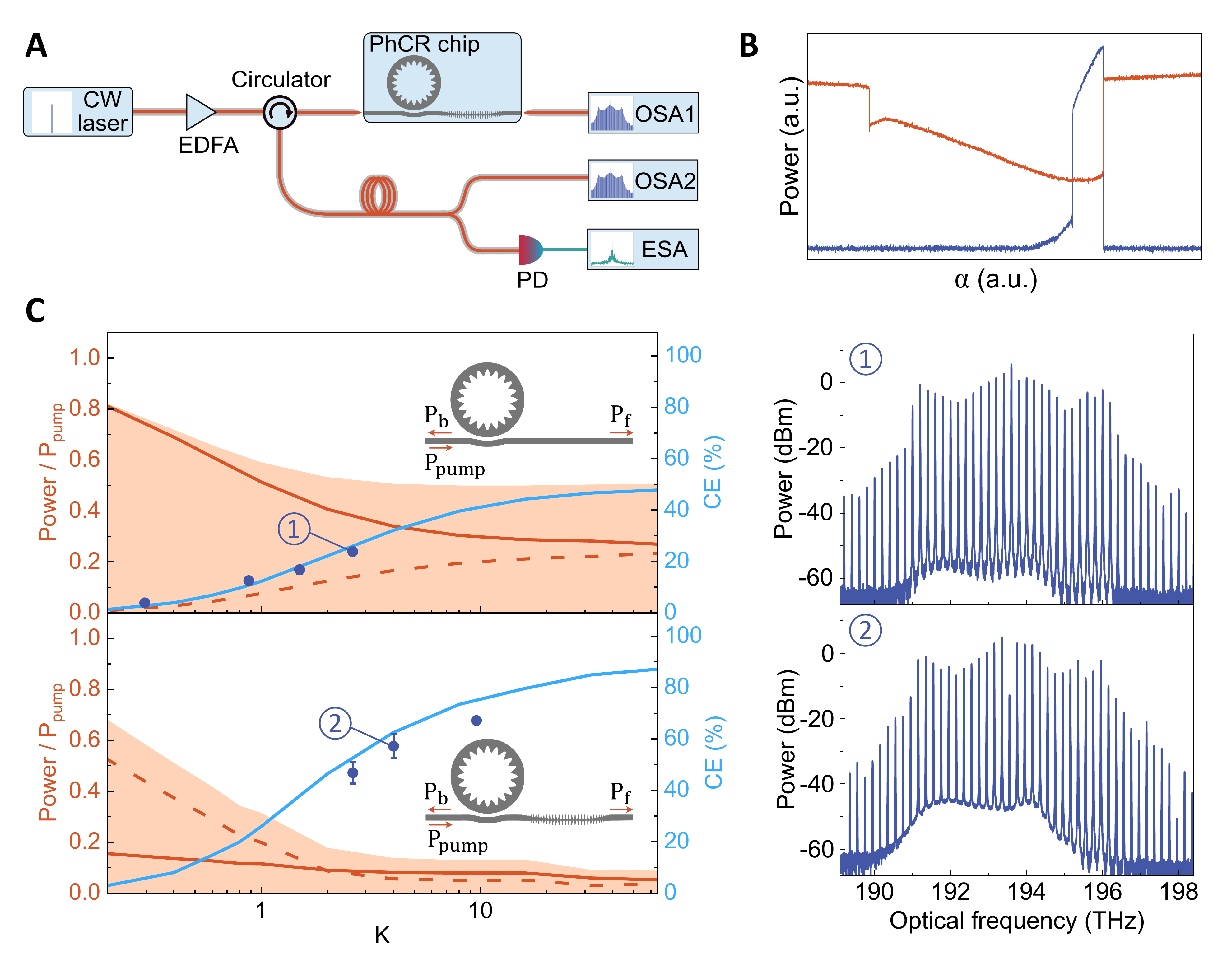}
\caption{ Laser-power consumption and CE in the nanophotonic circuit. (A) Experimental setup for soliton generation and characterization. EDFA, erbium-doped fiber amplifier; ESA, electrical spectrum analyzer. (B) Oscilloscope traces of $P_b$ (red) and $P_{\rm{comb}}^b$ (blue) versus $\alpha$. (C) Comparison of unused pump power and CE between the cases with $R=0$ (upper panel) and  $R=0.9$ (lower panel). The plot includes $P_f / P_{\rm{pump}}$ (solid red line), $P_b / P_{\rm{pump}}$ (dashed red line), $(P_{f}+P_{b})/ P_{\rm{pump}}$ (red shaded area), and CE (solid blue line) as a function of $K$ in LLE simulation. The blue circles represent the measured CE and the right panels shows the corresponding optical spectra of soliton microcombs. }
\label{Fig3}
\end{figure}

We explore the operation of our nanophotonic circuit by comparison to the case with $R=0$; see Fig. 3.  In both cases,  the tantala PhCRs we investigate have an intrinsic quality factor $Q_i=\omega_0/\kappa_i=2.7\times10^6 $. Other parameters of the devices include $a=109.5$ $\mu$m, RW = 4 $\mu$m, waveguide height = 570 nm, $A_{\rm{in}}$ = 275 nm, FSR = 200 GHz,  $D_2=-2\pi\times8.5$ MHz, and $\epsilon$ = 0.9 GHz. Our setup for soliton generation and characterization is shown in Fig. 3A; here we use an erbium-doped fiber amplifier (EDFA) to boost the power of our tunable laser. We characterize soliton generation by measuring at both the forward and backward ports. To verify the solitons to be stable and free of breathing oscillations, we photodetect a portion of the circuit output and monitor with an electrical spectrum analyzer (ESA). Figure 3B shows the primary, real-time signals of soliton formation in the case with a reflector and $\phi$ optimally set. As we scan $\alpha$, we monitor $P_b$ (red) and $P_{\rm{comb}}^b$ (blue). The soliton microcomb forms and stabilizes without an overly abrupt change in the intraresonator power, mitigating thermal bistability. Stabilizing a soliton with $R=0$ is marginally more sensitive to system fluctuations, such as those of the on-chip pump power and pump detuning.

With the capability to form and measure solitons with nearly arbitrary settings of $R$ and $\phi$, we present predictions of laser-power consumption and CE, emphasizing how the reflector closes off excess loss in the forward direction and enables near unit CE. We plot predictions for $P_f / P_{\rm{pump}}$ (solid red line), $P_b / P_{\rm{pump}}$ (dashed red line), and CE (solid blue line) as a function of $K$ in Fig. 3C. Here, the behavior with $K$ is important because power flows in our circuit through several channels: internal losses of the PhCR, coherent scattering to balance the bidrectional PhCR, and the forward and backward output ports. Moreover, we optimize $\alpha$ for maximum CE at each setting of $K$ and $P_{\rm{pump}}$. The ratio of unused pump power, $(P_{f}+P_{b})/P_{\rm{pump}}$, indicates laser consumption in our nanophotonic circuit as shown by the red shaded area in Fig. 3C. The limit of $K\gg1$, which we consider below, simplifies understanding because the internal losses can be neglected. 

Now, we directly compare the effect of the reflector. In the case with $R=0$ (upper panels of Fig. 3C), $P_b$ is loaded by conversion to the soliton and $P_f$, and it contains about 25\% of the pump for $K\gg1$. At most 50\% of the pump is converted to the soliton, indicating clamped CE from the excess output of double-sided emission. Indeed, our measurements of CE versus $K$ (blue circles in Fig. 3C) confirm the simulation. With the reflector set to $R=0.9$, $P_f$ is substantially reduced for all values of $K$; see the lower panels of Fig. 3C. Moreover, the backward pump inside the resonator is enhanced because we set $\phi$ for optimal constructive interference in both the measurements and the simulation. With appropriate $P_{\rm{pump}}$ and $\alpha$, this one-sided nanophotonic circuit enables conversion of the enhanced backward pump to the soliton microcomb, consuming in principle all of the input power. Moreover, the ratio of unused pump power exhibits a rapid decreases with $K$, reaching 0.2 at $K=2$. Even larger $K$ reduces the ratio of unused pump power to below 0.1 and increases CE to 87\%, which is ultimately limited by the intrinsic loss of the resonator and non-unit $R$ of the reflector. The panels labeled \circled{1} and \circled{2} at right in Fig. 3C show optical spectra of soliton microcombs that correspond to the CE measurements. In both cases, we observe an approximately constant power per mode: there are 20 comb modes with spectral power variation below 10 dB. Particularly, in Panel \circled{2}, the nanophotonic circuit can generate 14 comb lines with power above 0.5 mW at $P_{\rm{pump}}$ = 33 mW.  Moreover, the pump mode is 20 dB lower than the adjacent comb lines, indicating high laser-power consumption.

 \begin{figure}[htb]
\centering
\includegraphics[width=1\linewidth,trim={0.3cm 10.8cm 0.1cm 1.7cm},clip]{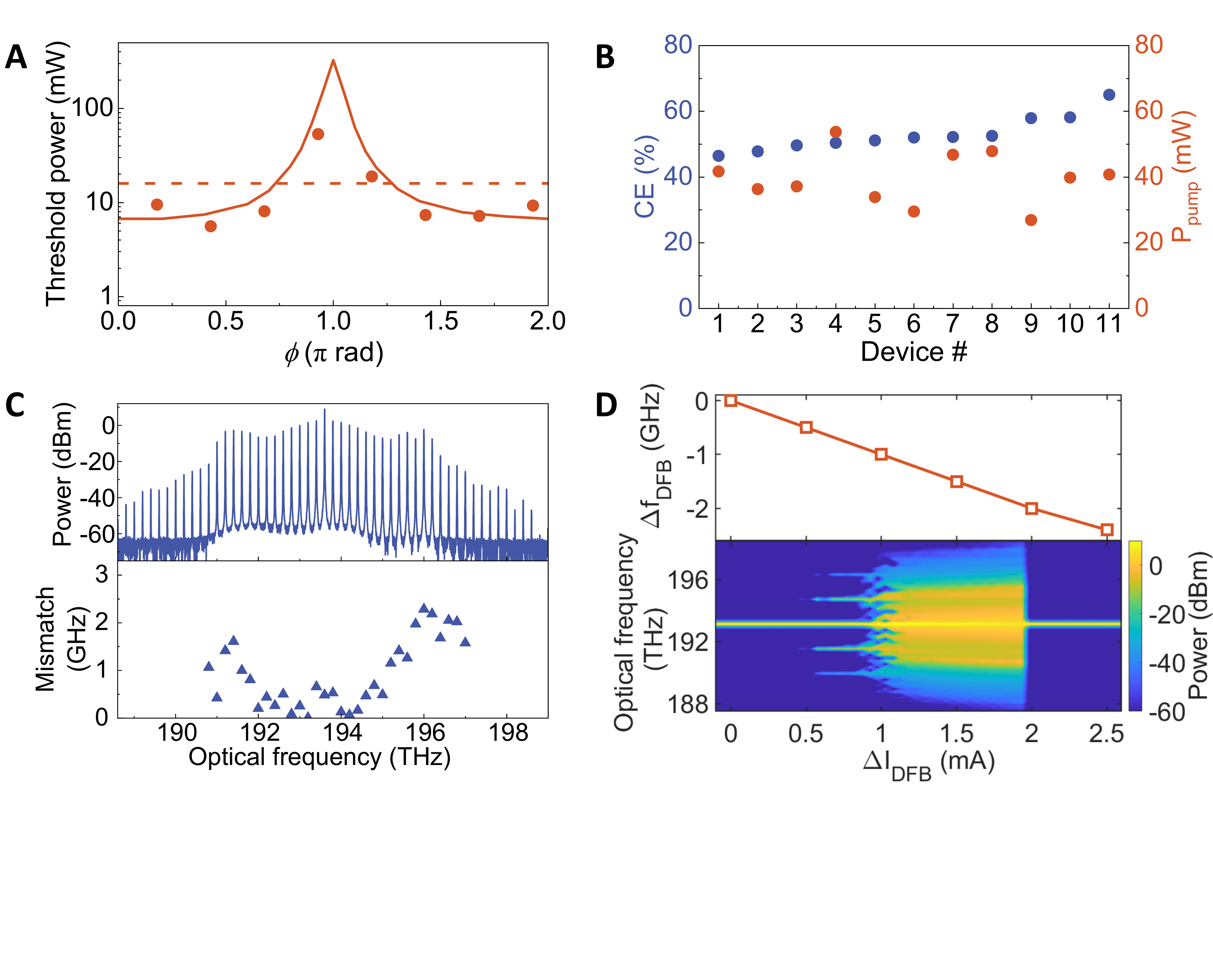}
\caption{ Practical characteristics of the nanophotonic circuit. (A) Comparison between the measured (red circles) and simulated (solid red line) threshold power versus $\phi$ in the nanophotonic circuit. The dashed red line shows that of a PhCR with the same parameters except $R=0$. (B) Measured CE (blue circles) and $P_{\rm{pump}}$ (red circles) of 11 nanophotonic circuit devices on a single chip. (C) The optical spectrum of a 200 GHz soliton microcomb (upper panel) and its frequency mismatch with ITU-T grid (lower panel). (D) Frequency tuning of the DFB laser from 193.1 THz (upper panel) and the associated spectrum generated by the nanophotonic circuit (lower panel) as we increase the laser current from 117 mA.}
\label{Fig4}
\end{figure}

Since our nanophotonic circuit generates a soliton microcomb with favorable properties for applications, we study some of its practical characteristics, such as operating power, device yield in fabrication, and  compatibility with a DFB laser diode as pump source; see Fig. 4. The reflector phase also influences the threshold power for optical parametric oscillation, the precursor of soliton formation, since an optimal setting of $\phi$ maximally enhances the intraresonator pump. Figure 4A shows measurements (red circles) and the LLE simulation (solid red line) of threshold power versus $\phi$ with the reflector. In this experiment, we undercouple the resonator with $K=0.5$ to emphasize the reduced threshold power in our nanophotonic circuit while other parameters of the devices are the same as that in Fig. 3. The threshold power for a PhCR without a reflector (dashed red line in Fig. 4A) is a factor of 2.9 higher than the optimal reflector circuit, since the pump power is bidirectionally distributed not only inside the PhCR but also in the coupling waveguide. 

The capability to yield a large fraction of fabricated nanophotonic circuits is an important consideration for applications. In particular, fabrication imprecision can degrade performance due to the sensitivity of $\phi$ and $K$ to device geometry. We study this by testing several devices to assess variation of CE and the $P_{\rm{pump}}$ setting used to operate the circuit; see Fig. 4B. Here, we select 11 nanophotonic circuit devices on a single chip; all of the devices that we fabricate are operational but the designed parameter variations lead to reduced yield. Among the devices, CE varies around 50\% and $P_{\rm{pump}}$ needed for operation varies around 40 mW.  These values are consistent with our overall observed tolerance, and we optimize the fabrication procedure and scanning electron microscopy inspection to maintain a consistent $K$.

Beyond CE, microcomb applications also benefit from frequency alignment to the ITU-T grid, the international channel standard for data communication systems. Such alignment ensures compatibility with system components like wavelength-division multiplexers and demultiplexers \cite{Kiyoul2022}. We demonstrate robust frequency alignment of the soliton microcomb generated in our nanophotonic circuit. In Fig. 4C, we show the optical spectrum of a soliton microcomb with a precise 200 GHz mode spacing. By varying $a$ and RW of the PhCR, we finely tune both FSR towards a target of 200 GHz and $\omega_0$ to align with any pre-defined grid. We present measurements of the frequency mismatch in Fig. 4C in which the 27 microcomb modes between 191 THz and 197 THz are aligned to the ITU-T grid better than 2.6 GHz. 

To demonstrate the compatibility of our nanophotonic circuit with chip-scale lasers, we use a standard, butterfly-packaged DFB laser as the pump; see Fig. 4D. Such DFB lasers offer frequency adjustment through the control of laser temperature and current; in particular, the frequency-control parameters of our laser are -11.3 GHz/K and -0.9 GHz/mA, respectively. Hence, we tune the DFB laser temperature to coarsely align the laser with the split mode, and then we use the DFB laser current to sweep $\alpha$ for soliton generation. Figure 4D illustrates the DFB laser frequency change ($\Delta f_{\rm{DFB}}$) with current and the associated spectrum generated by the nanophotonic circuit. We observe optical parametric oscillation followed by the evolution to a stable soliton. This straightforward and reliable technique combined with low required pump power and high CE of our nanophotonic circuits points the way to a scalable, frequency-comb laser source for high-capacity telecommunication applications.

In conclusion, we have presented a nanophotonic resonator circuit that consists of a bidirectional Kerr resonator and a reflector on the coupling waveguide. Coherent backscattering with nanophotonics enables arbitrary phase-matching and high laser-power consumption in the normal dispersion PhCRs, allowing for soliton formation at approaching unit CE. Using this nanophotonic circuit, we generate 200 GHz flattop solitons with up to 65\% CE and 97\% pump consumption. We investigate the nonlinear dynamics and steady-state fields in the nanophotonic circuit with LLE and observe agreement between simulation and measurement results. Moreover, we study some practical characteristics of the nanophotonic circuit by demonstrating high yield fabrication of the high-efficiency devices, generation of comb lines that are precisely aligned with ITU-T frequency grids and the compatibility with a chip-scale DFB pump laser for future integration \cite{ChinmayOL,Bergman2023}.

We thank Atasi Dan, Yang Li, and Daniel Slichter for reading the manuscript. 
This work is a contribution of NIST and not subject to US copyright. This research has been funded by DARPA PIPES, AFOSR FA9550-20-1-0004 Project Number 19RT1019, NSF Quantum Leap Challenge Institute Award OMA - 2016244, and NIST on a Chip. Mention of specific companies or trade names is for scientific communication only and does not constitute an endorsement by NIST.

\bibliography{scibib}
\bibliographystyle{Science}
\end{document}